\documentclass{sig-alternate}
\usepackage{german,times,amsmath,amssymb,epsfig,url,algorithm,algorithmic}

\begin{document}

\setlength{\pdfpageheight}{\paperheight}
\setlength{\pdfpagewidth}{\paperwidth}

% You may need to change the horizontal offset to do what you
% want.  Setting \hoffset to a negative value moves all printed
% material to the left on all pages; setting it to a positive value
% moves all printed material to the right on all pages; not setting
% it keeps all printed material in it's default position.  \voffset
% is the vertical offset: use negative value for up; don't set if
% you want to use default position; use positive for down.
% \hoffset = -0.2truein
% \voffset = -0.2truein

% --- Author Metadata here ---
\conferenceinfo{} {}
\CopyrightYear{2011}
\crdata{978-1-4503-0493-1/11/02}
\clubpenalty=10000
\widowpenalty = 10000

\newcommand{\squishlist}{
 \begin{list}{$\bullet$}
  { \setlength{\itemsep}{0pt}
     \setlength{\parsep}{3pt}
     \setlength{\topsep}{3pt}
     \setlength{\partopsep}{0pt}
     \setlength{\leftmargin}{1.5em}
     \setlength{\labelwidth}{1em}
     \setlength{\labelsep}{0.5em} } }

\newcommand{\squishlisttwo}{
 \begin{list}{$\bullet$}
  { \setlength{\itemsep}{0pt}
     \setlength{\parsep}{0pt}
    \setlength{\topsep}{0pt}
    \setlength{\partopsep}{0pt}
    \setlength{\leftmargin}{2em}
    \setlength{\labelwidth}{1.5em}
    \setlength{\labelsep}{0.5em} } }

\newcommand{\squishend}{\end{list}}

\newtheorem{theorem}{Theorem}
\newtheorem{lemma}{Lemma}
\newtheorem{corollary}{Corollary}
\newtheorem{definition}{Definition}
\newcommand{\entity}[1]{\emph{#1}}
\newcommand{\newWord}[1]{\emph{#1}}
\newcommand{\ignore}[1]{}
\newcommand{\comment}[1]{}
\newcommand{\indented}[1]{
	\begin{tabbing}
		blah\=\+\kill
		\parbox{8cm}{#1}
	\end{tabbing}
}
\newcommand{\ruleline}[1]{\indented{#1}}

\title{An Operator for Entity Extraction in MapReduce}

%\subtitle{[Extended Abstract]
%\titlenote{A full version of this paper is available as
%\textit{Author's Guide to Preparing ACM SIG Proceedings Using
%\LaTeX$2_\epsilon$\ and BibTeX} at

\numberofauthors{1}
\author{Ndapandula Nakashole\\
  Carnegie Mellon University \\
  5000 Forbes Avenue \\
  Pittsburgh, PA, 15213 \\
  {\tt ndapa@cs.cmu.edu} 
}

\comment{
\numberofauthors{4}
\author{
% 1st. author

% 2nd. author
}
} %\comment

\maketitle

\begin{abstract}
Dictionary-based entity extraction involves finding mentions of
dictionary entities in text. Text mentions are often noisy, containing spurious or missing words.
Efficient algorithms for detecting approximate entity mentions follow one of  two general 
techniques. The first approach is to build an \textit{index} on the entities and perform
index lookups of document substrings. The second approach recognizes that the number 
of substrings generated from documents can explode to large numbers, to get around this,
they use a \textit{filter} to prune  many such substrings which do not match any dictionary
entity and then only verify  the remaining substrings if they are entity mentions of dictionary
entities, by means of a text join. The choice  between the \textit{index-based} approach and the
\textit{filter \& verification-based}  approach is a case-to-case decision as the best approach
depends on the characteristics of the input entity dictionary, for example frequency of entity
mentions. Choosing the right approach for the setting can make a substantial difference
in execution time. Making this choice is however non-trivial as there are parameters
within each of the approaches that make the space of possible approaches very large.
In this paper, we present a cost-based operator for making the choice among execution plans for entity extraction.
Since we need to deal with large dictionaries and even larger large datasets, our operator is developed
for implementations of  MapReduce distributed algorithms.

%Our experiments show that our operator-centric approach results in significant gains in execution time.

\end{abstract}


\section{Introduction}
Dictionary-based entity mention extraction has wide use in search and semantic web related work. For example, shopping portals annotate text documents with dictionary products to maximize product relevance to user queries. Semantic search detects mentions of entities such as people, organizations and locations in text documents in order to facilitate entity-oriented search.

% mentions are noisy
Determining wether a  document mentions a product is a challenging task, particulary becuase entity mentions in Web documents are noisy. Such mentions are rarely exact matches of the dictionary entities, which can be too long for users to write in full everytime they refer to the entity. For example, in product reviews, it is common for reviewers to use short representations of the entities in product catalogs. A mention in a document may miss some of the words of the entity or it may have extra words not found in the entity name in the dictionary,  therefore it is crucial for algorithms  to detect \textit{approximate mentions} of entities in addition to exact mentions \cite{AgrawalAK10, ChaudhuriGX09,CohenS04}

%  finding mentions 
Finding entity mentions entails finding substrings in a document sequence such that the substring matches an entity in the dictionary.
Such  \textit{candidate substrings} ,  are  substrings whose words are a full or partial subset of the words of an entity in the dictionary. A naive method to generate \textit{candidate substrings} would be to scan a document, generating all substrings of up to size $L$ where $L$ is the longest dictionary entity and do a dictionary lookup for each of the substrings.  Efficient algorithms for this problem follow one of two approaches. The first approach is to build an \textit{index} on the entities and perform index lookups of document substrings \cite{AgrawalAK10, ChandelNS06}. The second approach recognizes that the number of substrings generated from documents can explode to large numbers, to get around this, they use a \textit{filter} to prune  many such substrings which do not match any dictionary entity \cite{ChakrabartiCGX08} \cite{ChaudhuriGX09} and then only verify  the remaining substrings if they are entity mentions of dictionary entities, by means of a text join. If the longest entity in the dictionary has \textit{L} words. Given a document $d$, all substrings with length up to \textit{L} are possible mention candidates. This produces $L \times |d|$ substrings to be looked up. The filter serves to reduce the number of lookups as only substrings that pass the filter are verified if they in fact refer to a dictionary entity or not.

% avaiable choice, introduce options,  indexing schemes, 
The choice  between the \textit{index-based} approach and the \textit{filter \& verification-based} approach is a case-to-case decision as the best approach depends on the characteristics of the input entity dictionary, for example frequency of entity mentions. For non-distributed algorithms, the main differentiating factor is that the filter is a more compact structure than the index, in most cases the filter fits in memory whereas the index does not.  Performace of the  index-based approach is  affected by  the indexing scheme. Indexing on single words has a different effect to, for example, indexing sets of words that frequently co-occur in  dictionary entities. Individual word index posting lists can grow too long which incur long times for merging posting lists. Performance of the filter \& verification-based approach is affected by the type of filter used to prune substrings, a basic filter like a prefix-based filter for pruning out prefixes not likely to match dictionary entities would perform differently from a probabilistic filter such a Latent Signature Hash (LSH) filter.

If we consider the indexing scheme to be a paremeter and the filter to be a paremeter, than the number of available approaches can be  large and the choice becomes a challenge. In this paper we propose to add to this space of approximate entity mention extraction methods. Our proposal is based on the fact that while performance of the input entity dictionary at hand, the dictionary can be heterogenous such that it consists of partitions that if each partition is processed seperately by the most suitable method, it  can result in cheaper aggregate run times  than any of the methods  applied to the entire dictionary. This is the hybrid approach. Given the choices of indexing schemes and filters, and the hybrid approach combining index-based and fiilter \&verification based methods, the important problem we address is this paper is that of which of the indexing schemes to use, which of the filters to use and wether or not to use a hybrid approach, and if using a hybrid approach where should we partition the entity dictionary. This is an optimization problem over a large space of approaches.

% Need for MapReduce
We solve this optimization problem over distributed algorithms, since we need to deal with large dictionaries and even larger large datasets, we develop and optimize over scalable algorithms in the form of MapReduce distributed processing. In a MapReduce setting, network time plays a significant role on job completion time, furthermore, additional  MapReduce-specific coordination tasks such as disk-based sorting introduce costs that are not part of single machine algorithms. Additional distributed setting variables make the choice between the \textit{index-based} approach and the \textit{filter \& verification-based} approach more challenging. To differentiate between these parallel computation coordination tasks and actual processing time spent on the job, we make a distinction between \textit{work done time} and  \textit{ job completion time} in the objective functions we define for optimization.

In this paper, we introduce an operator named the \textit{Entity Extraction Join Operator,  EE-Join Operator}  for optimizing  entity extraction in MapReduce.  We define  the space of  approaches for approximate entity mention extraction, and propose a hybrid approach. We develop a cost model for estimating the costs of each of the approaches and  efficiently search the space of available options.

In summary we make the following contributions:

\begin{itemize}
\item An operator, \textit{EE-Join} for highly scalable and optimized entity extraction in MapReduce, which works with different objective functions, 
we use two distinct objective functions, the \textit{work done time} and \textit{job completion time}
\item A study  the space of  available approaches for approximate entity mention extraction, and propose an additional approach
\item A cost model for estimating execution time for the twor objective functions.
\item A means to gather data statistics needed leveraged  by the cost model.
\item An efficient algorithm for searching the space of available approaches.
\item An experimental evaluation of our operator, with  entity dictionaries consisting of entities that follow
various mention distributions.
\end{itemize}

The rest of this paper is organized as follows.  We next give the semantics of dictionary entity extraction in Section 2. Section
3 describes approximate entity mention algorithms and presents our adaptations of single machine algorithms
to their MapReduce counterparts that form the building blocks of the Entity Extraction Join operator. Section 4 describes the costmodel. Section 5 describes the optimization problem over the cost model as the objective function and presents our solution to the optimization problem.
 We report our experimental results in Section 6. Section 7 is a review of related work on optimization in text analytics and joins in MapReduce. Finally, we conclude in Section 8.

\section{Semantics of  Dictionary Entity Extraction}
In this section, we formally define dictionary-based entity extraction, our focus is on approximate mentions where substrings can be partial matches of the dictionary entitites. The similarity function used plays an important role in the semantics of approximate mentions, we therefore define and motivate similarity functions we use.
 
Approximate dictionary-based entity extraction  takes as input a dictionary of entities, $\xi$ and a document collection $\zeta$ to output all pairs $(e, d)$, such that $sim(e, d_{s(i)})>= \gamma$ where $e \in \xi$, $d_{s(i)}$ is a substring in $d$, $d \in \zeta$ and  $\gamma$ is a given similarity threshold.

To compute pair-wise similarity $sim(e, d_{s(i)})$, different similarity functions can be used depending on the desired semantics.
% why not jaccard similarity  and why Jaccard containment
A commonly used similarity measure is  \textit{Jaccard similarity} \cite{HadjieleftheriouL09, KoudasSS06},  defined as: $JaccSim(e, d_{s(i)})$ = $\frac{e \cap d_{s(i)} } { e \cup d_{s(i)} }$.  Jaccard similarity is a symmetric measure; where asymmetric semantics are desired a different measure is needed. For example, suppose the dictionary contains two entries,
\textit{E1:iPhone Charger} and \textit{E2: Apple iPhone 4 Black or White 32G AT\&T}. Assuming a document contains a  substring \textit{S1: iPhone 4}. Then $JaccSim(E1, S1)$ = $\frac{1}{3}$, but $JaccSim(E2, S1)$ = $\frac{1}{4}$, although \textit{E2} is semantically the better match. Jaccard Similarity gives a lower score to \textit{E2} because a  large fraction of its words are missing in \textit{S1}. A measure that reflects the fact that \textit{S1} is fully contained in \textit{E2} would solve this problem. Prior work has defined an asymmetric measure \textit{Jaccard containment} as: $JaccCont(e, d_{s(i)})$ = $\frac{e \cap d_{s(i)} } { d_{s(i)}}$.\\ Thus the Jaccard containment of \textit{iPhone 4} in \textit{E1}, $JaccCont(E1, S1)$ = $\frac{1}{3}$, whereas  $JaccCont(E2, S1)$ = $1$.

We  use Jaccard containment as the similarity measure. We note that there are two important variations for the  \textit{Jaccard containment } measure. The first Jaccard containment variation tolerates missing words in the  approximate mention, as in the example above. The second variation tolerates extra tokens in  the approximate mention.  Clearly, the semantics fo the Jaccard containment variations
are different and one may be desirable in settings where the other is not. % give example where this may be the case
\\

\noindent {\bf DEFINITION 1:}
Given an entity $e$ and a document substring $d_{s(i)}$, the Jaccard containment of $d_{s(i)}$  in  $e$, allowing for  \textit{missing} words in $d_{s(i)}$, is: $JaccCont^{missing}(e, d_{s(i)})$ = $\frac{e \cap d_{s(i)} } {d_{s(i)}}$. Allowing \textit{extra} words in $d_{s(i)}$, $JaccCont^{extra}(e, d_{s(i)})$ = $\frac{e \cap d_{s(i)} } {e}$.
\\

Using Jaccard containment and its variations we can leverage an interesting property that enables efficient computation of  Jaccard variants. Given a similarity threshold $\gamma $, we can compute all substrings $s_{ij} \subseteq s$ such that the total weight of words  in $s_{ij}$, $w(s_{ij})$ is $ \geq w \times \gamma $.  All these substrings are  approximate mentions according to Jaccard containment, we refer to these as the the Jaccard variants of a string.  For example,  consider the entity, \textit{Apple iPhone 4 32G}. Suppose the token weights are as follows: \textit{Apple:\{1\}, iPhone:\{8\}, 4:\{2\}, 32G:\{1\}}. For $\gamma  = 0.75$,  the Jaccard variants is of the entity are:
\{Apple iPhone 4\}, \{iPhone 4\}, \{iPhone 4 32G\}, \{Apple iPhone 4 32G\}.
If we store compute and store all Jaccard variants of dictionary entities, we can perform exact match comparisons between the Jaccard variants of the dictionary entities and the Jaccard variants of potential mentions. 
\\

\noindent {\bf DEFINITION 2:}
A subsequence $s_{ij} \subseteq s $  is  a Jaccard variant of $s$ if $s_{ij} \cap {s} >= \gamma $.
For settings where words are \textit{weighted}, a weighted subsequence $ws_{ij} \subseteq ws$ is a Jaccard variant of $ws$ whose weight is $w$, if  $w(s_{ij} \cap {s}) >= w \times \gamma $.
\\

% Discussion of how to compute.
Computing the Jaccard variants for the dictionary entities is straightforward, however computing the 
the Jaccard variants if done naively can explode since every substring  in a document is potentially a Jaccard variant
of some dictionary entity.  All these have to be queried against the variants of the dictionary entities. We avoid generating all possible Jaccard variants as explained later.
% in Section ??.

\begin{figure}[t]
%\vspace*{1cm}

\begin{center}
\includegraphics[scale=0.98,bb= 0 0 160 160]{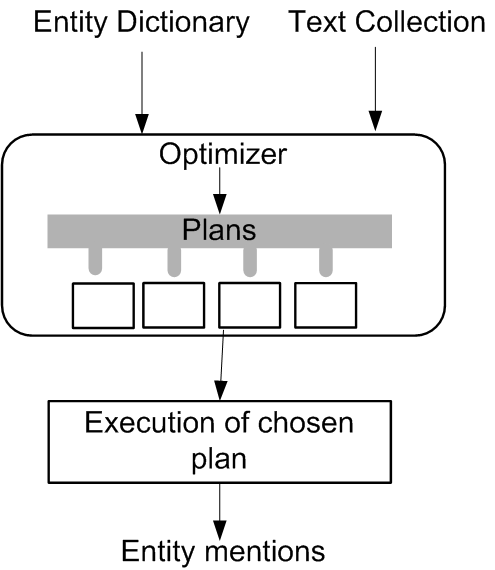}
%\includegraphics[width=1\linewidth]  {overview.pdf}
%
%\vspace*{-1cm}
\caption{Overview of the EE-Join operator.}
\label{fig:eejoin}
\end{center}
\end{figure}

%\begin{itemize}
%\item Jaccard similarity.
%
%\item Jaccard containment (extra)
%
%\item Jaccard containment (missing)
%
%\item Transformations
%\end{itemize}

\section{Approximate Mention Algorithms in MapReduce}
Having defined the semantics, we can introduce algorithms for approximate entity mention extraction. For each 
algorithm we briefly describe the single machine version before explaining how adaption to MapReduce is realized
through the use of MapReduce constructs. We then explain the impact of the MapReduce constructs on the performance for each of the algorithms.

\subsection{MapReduce SSJoin}
Chaudhuri et al. \cite{ChaudhuriGK06}  introduced the notion of set similarity join (SSJoin) for identifying similar strings. They observe that typically efficient algorithms  for similartity joins use a  similarity function chosen to suit the domain and application.  The premise of SSJoin is to decouple similarity functions from the implementation of similarity joins, instead they propose an operator  as a foundation to implement similarity joins that can adaptively handle a variety of  similarity functions. For example, depending on the size of
the relations being joined and the availability of indexes, the SSJoin optimizer may choose either index-based plans or merge and hash joins in order to implement the SSJoin operator. 

The simplest implementation of the SSJoin on a single machine thus compares every substring $s$ to every entity $e \in \xi$.
We adapt the SSJoin algorithm to MapReduce to create  a baseline MapReduce algorithm. The mappers generate all substrings $s$ of length $l$ (maximum entity length) for every document $d \in \zeta$.  The mappers then generate one or more signatures for each of the substrings. The same mapper functions are applied to the entire dictionary such that for each $e \in \xi$ , the mappes generate signatures, applying the same signature generating function.  The signature generating function is constructed such that if a substring and a dictionary entity are similar, they will at least have one signature in common.  Thus, using  signatures as the reduce key, substrings and entities with a signature in common are shuffled to at least one common reducer which computes similarities between substrings. One of the shortcomings of this algoirthm is that it requires a reduce function. This incurs a significant amount of data transfer  time required to shuffle the substrings and entities to  the reducers.  The shuffling cost of this baseline algorithm is exacerbated by the fact that all possible substrings from all documents are generated.
The MapReduce algorithm is outlined in  Figure \ref{fig:baselineSSJ}. 

\begin{figure}[!t]%
\begin{minipage}[t]{1\linewidth}
\hrule
\begin{tabbing}
\textbf{fun}\=\textbf{ction} map($item_{i}$, $tag$) \\

\> \textbf{if} \=  $item_{i}$ is a document \\
\> \> List ${S} \leftarrow$ generate all substrings from document ($d_{i}$) \\
\> \textbf{el}\=\textbf{se if} $item_{i}$ is an entity \\
\> \> List ${S} \leftarrow$  $item_{i}$ \\

\> \textbf{fo}\=\textbf{r}   ${s}$ $\in$ $S$ \textbf{do}\\
\> \> generate signatures from $s$ \\
\> \> emit signatures \\\\

\textbf{fun}\=\textbf{ction} reduce($sig_{i},[item_{1, entity}, item_{12, entity}, item_{31,doc}, ...]$) \\
\> Hashtable $MatchingEntities$  $\leftarrow$ \{ \} \\
\> \textbf{if} \=  $item_{i}$ is an entity \\
\> \> add $item_{i, entity}$ to hashtable $MatchingEntities$  \\

\> \textbf{el}\=\textbf{se}  \\
\> \> \textbf{fo}\=\textbf{r}   $item_{i, entity} \in MatchingEntities$ \textbf{do}\\
\> \> \> \textbf{if} \=  $sim(item_{i, entity}, item_{j, doc}) \geq \gamma$  \\
\> \> \> \> emit$(item_{i, entity}, item_{j, doc}$)
\end{tabbing}
\hrule
\end{minipage}%
\caption{MapReduce \textit{SSJoin} baseline}
\label{fig:baselineSSJ}
\end{figure}

\subsection{MapReduce Index on Entities}
The index on entities approach creates an index on words of the dictionary entities.
It then generates all substrings and queries the index for similar entities.
We adapt the index-based approach to entities as follows: First we generate
the index on the entities as a seperate MapReduce job. The index is then broadcast
to every Mapper node. The dictionary is broadcast to every Mapper node. The type of index
used can vary from application domain. Assuming a basic index with inverted lists per word, and then for each query substring $q$
retrieve all lists corresponding to words in  $q$. The union of the lists is the candidates entities
that are mentioned by $q$. Each of the candidates entities are then verified to determine if
they are true mentions. The MapReduce algorithm is outlined in  Figure \ref{fig:indexOnEntities}. Though the index is created in separate MapReduce it is not a significant portion of the execution time as the dictionary is typically much smaller than the document collection.
One of the limiting factors of this algorithm is the fact that the index can be large, may not fit in memory. This means the index has to be partitioned into smaller indices which can fit in memory, and the entire corpus has to be processed serveral times, once for every index partition.
The MapReduce algorithm is outlined in  Figure \ref{fig:indexOnEntities}. 

The type of index used plays an important role in the performance of the algorithm. We studied
three types of indices and their properties.

\begin{itemize}
\item  \textbf{Per word index}:  This is the basic index, where an inverted list is generated for
every word, storing all entities consisting of that word.  While single word inverted index can be generated quickly, 
these lists can grow very large, making the task of list merging expensive.

\item \textbf{Prefix-index}: A prefix index arranges the words of the entities according to a fixed order, for example, 
based on decreasing occurrence frequency. During similarity join, we have to generate the prefixes of the substrings and query them against
the index to verify that the substring-entity match surpasses a user specified threshold.  Like the per word index, the prefix index is quick to generate, its advatange is that it reduces the problem of potentially long inverted lists.

\item \textbf{Jaccard variant index}:  The Jaccard variant index is an index on all the Jaccard variants of all the entities.
During similarity join, we have to generate the Jaccard variants of the the substrings. The advantage of the Jaccard Variant index is
that it requires no verification, a substring with a Jaccard variant with an inverted list $il$ is an approximate mention
of all those entities with the Jaccard variant as a substring. Constructing a Jaccard vairant index is slightly more expensive then other two.
\end{itemize}

\begin{figure}[!t]%
\begin{minipage}[t]{1\linewidth}
\hrule
\begin{tabbing}
\textbf{fun}\=\textbf{ction} map($index_{P}$, $d_{i}$) \\
\> Index ${Idx} \leftarrow$ load index into memory\\

\>List ${S} \leftarrow$ generate all substrings from document ($d_{i}$) \\
\> \textbf{fo}\=\textbf{r}   ${s}$ $\in$ $S$ \textbf{do}\\
\> \> lookup $s$  on the index\\
\> \> \textbf{if} \=  $sim(e, s) \geq \gamma$  \\
\> \> \> emit$(e, s$)
\end{tabbing}
\hrule
\end{minipage}%
\caption{MapReduce \textit{Index-based} lookups}
\label{fig:indexOnEntities}
\end{figure}

\subsection{MapReduce ISHFilter \& SSJoin}

We have introduced the baseline SSJoin algorithm and the index-based algorithm, both generate all
possible substrings and then computing a similarity join between a large set of substrings and the dictionary entities. 
Perfoming a similarity join between all substrings is a large peformance bottleneck. To overcome this problem we first filter out all substrings
that cannot match with any dictionary entity, and only then, perform a set similarity join (SSJoin). We use the 
The ISHFilter introduced by  Chakrabarti et al.  \cite{ChakrabartiCGX08} to prune a large number of substrings that are obvious
non-mentions.  The SSJoin algorithm is then applied to remaining substrings with the dictionary entities.
The difference between baseline \textit{SSJoin} and \textit{ISHFilter \& SSJoin} is that shuffling cost is much
lower as number of substrings is substantially reduced by the filter.
The MapReduce algorithm is outlined in  Figure \ref{fig:FilterSSJ}. 

So far we have introduced the SSJoin as using a signature scheme to generate sinatures such that substrings and entities with a signature in common are shuffled to atleast one reducer. The signature used upon which the data is shuffled plays a significant role in performance. 
We studied three signature schemes.

\begin{itemize}
\item \textbf{Single word signatures}:  Using each word as a signature has a lot of skew, becuase some words are very common.  This results in high shuffling costs. Since each substring and entity consists of many words, this type of signature results in duplicate work at the reducers.  

\item \textbf{Prefix signatures}:   The prefix signature uses prefixes as signatures. While quick to generate, the
prefix signatures are susceptible to skew, causing shuffling costs to be skewed to a few nodes. The prefix signature requires
verification at the reducers.

\item \textbf{Locality-Sensitive Hashing (LSH) signatures}:  is an algorithm for solving  approximate or exact similarity.  It is probabilistic in the sense that it uses a  hash on the the input so that similar items are mapped into the same group with high probability. Like Prefix signatures, LSH signatures require verification. 

\item \textbf{Jaccard variant signatures} : Jaccard variants as signatures has the advantage that it reduces  data skew, while also not requiring verification at the reducers.
\end{itemize}

\begin{figure}[!t]%
\begin{minipage}[t]{1\linewidth}
\hrule
\begin{tabbing}
\textbf{fun}\=\textbf{ction} map($item_{i}$, $tag$) \\

\> \textbf{if} \=  $item_{i}$ is a document \\
\> \> List ${U\_S} \leftarrow$ generate all substrings from document ($d_{i}$) \\
\> \> List ${S} \leftarrow$ apply \textit{ISHFilter} to all substrings $U\_S$ \\
\> \textbf{el}\=\textbf{se if} $item_{i}$ is an entity \\
\> \> List ${S} \leftarrow$  $item_{i}$ \\

\> \textbf{fo}\=\textbf{r}   ${s}$ $\in$ $S$ \textbf{do}\\
\> \> generate signatures from $ s{s}$ \\
\> \> emit signatures \\\\

\textbf{fun}\=\textbf{ction} reduce($sig_{i},[item_{1, entity}, item_{12, entity}, item_{31,doc}, ...]$) \\
\> Hashtable $MatchingEntities$  $\leftarrow$ \{ \} \\
\> \textbf{if} \=  $item_{i}$ is an entity \\
\> \> add $item_{i, entity}$ to hashtable $MatchingEntities$  \\

\> \textbf{el}\=\textbf{se}  \\
\> \> \textbf{fo}\=\textbf{r}   $item_{i, entity} \in MatchingEntities$ \textbf{do}\\
\> \> \> \textbf{if} \=  $sim(item_{i, entity}, item_{j, doc}) \geq \gamma$  \\
\> \> \> \> emit$(item_{i, entity}, item_{j, doc}$)
\end{tabbing}
\hrule
\end{minipage}%
\caption{MapReduce \textit{ISHFilter \& SSJoin} }
\label{fig:FilterSSJ}
\end{figure}

\subsection{MapReduce Index on Documents}
The fourth  approach  for approximate entity mention extraction on MapReduce is to create an index on the entire document collection. The input to the mappers are partitions of the document index, the dictionary of entities is broadcast to every mapper. Each dictionary entity $e \in \xi$ is treated as a query which is posed on the index. Each mapper searches its part of the document index for all entity queries. The complete list of mentions is the union of mentions found by each mapper. When the dictionary of entities does not fit in memory, the algorithm makes multiple passes over the document index. Constructing an index on the entire corpus is an expensive operation, this is unlike the approach that constructs the index on the dictionary of entities becuase the dictionary of entities is usually orders of maginitude smaller than the document collection. 
Since the index on the document collection is usually not available upfront, and constructing it is expensive, we do not persue this approach further in the rest of the study.

\subsection{  Operator Algorithms}
We have aldready eliminated the \textit{ Index on Documents} algorithm from the algorithms considered for the EEJoin operator.
We further eliminate the \textit{MapReduce SSJoin} algorithm due to its limitation with generating all substrings.  Instead we keep the optimized version of SSJoin, the \textit{MapReduce ISHFilter \& SSJoin}. We also keep the \textit{MapReduce Index on Entities} but instead of generating all substrings from documents, we use the filter to effectively have the \textit{MapReduce ISHFilter \& Index on Entities} approach. These two algorithms provide a rich set of options as both the allow tuning of the signatures used by the SSJoin and the type of index used by the entity indexing algorithm.

\section{Cost Models}
Having described the algorithms for the EEJoin operator, we now present the cost-model used to 
estimate performance of the each of the algorithms. Using  a cost model, we can automatically determine which of the algorithms has
the best performance for a given input dictionary $\xi$ and document collection $\zeta$. We consider two objective functions, one for the total work done and another for the  job completion time.

%\textit{work done vs total job completion time}
The job completion time  of the \textit{Index on Entities} approach is made up of two 
main componets. The first is the substring lookup time denoted by $C_{lookup}$ in Definition 3. The total lookup time is equally distributed among the  mappers due to the MapReduce load balancers for the mapper, thus total lookup time is $\left(\frac{|C|}{|M|} \times C_{lookup}\right)$.
The second is the number of iterations made over all the substrings due to th entity dictionary not fitting in memory, denoted by $ \frac{|E|}{M_{e}} $. 
\textit{explain estimation of |C| and |M| from data statistics.}
\\

\noindent {\bf DEFINITION 3:}
The cost of the index approach , for job  completion time, is defined as follows:
$Cost^{index} = \left(\frac{|C|}{|M|} \times C_{lookup}\right) \times \frac{|E|}{M_{e}}$
Where $|C|$ is the number of candidate substrings from all documents in the dataset, $|M|$ is the number of mappers,
$|E|$ is the dictionary size and $|M_{e}|$ is the memory budget for the index. Thus $\frac{|E|}{|M_{e}|}$ is the total number of passes made over the data.
\\

The job completion time of the \textit{ISHFilter \& SSJoin}  approach consists of  three main components. The first is the cost of generating signatures over $ \frac{|C|}{|M|}  \times C_{sig(c)}$ this cost depends on the type of signature used. The second is the cost  of shuffling the signatures over the network,  $(C_{suffle(sig)}$.  This cost depends  on the number of  signatures per candidate. The third is the cost of verifying the candidates of a signature, $C_{verify{sig}}$ as  shown in Definition 4.
\\

\noindent {\bf DEFINITION 4:}
The cost of the filter \& ssjoin approach, for job  completion time, is defined as:
$Cost^{ishf\&ssj} = \frac{|C|}{|M|}  \times C_{sig(c)}+ |Sig| \left(C_{suffle(sig)} + C_{verify{sig}} \right)$. Where
$C_{sig(c)}$ is the cost of generating signatures, $C_{suffle(sig)}$ is the cost of shuffling signatures and $ C_{verify{sig}}$ is the cost of verifying candidates of a signature.
\\

%\textit{Here explain estimation}

\section{Optimization}
We optimize over a large plan space of algorithms. The plan space is made up of two core algorithms. However, each of the algorithms can be instantiated with several signature schemes. Furthermore, the dictionary is partitioned such that a fraction of the entities is processed by one of the core approaches and the rest are processed by another core approach. Furthermore, any combinations of the different signature schemes form a possible hybrid approach, thus creating a large space of possible plans.
%
%\noindent {\bf DEFINITION 6:}
%A hybrid is defined to be a combination of the index and the filter \& SSJoin. Where the cut-off is specified.

\subsection{Plan space}
The \textit{EE-operator} optimizes entity extraction by partitioning entities into mention frequency
categories. The intuition is that each of the algorithms performs better for entities of certain mention
frequencies than the other approach. Therefore, for a given input dictionary and text collection, a fraction of
the entities is processed by the \textit{index-based} approach, for some  signature scheme, and the remaining fraction is processed by the \textit{filter \& verification-based} approach, for some signature scheme, not necessarily the same as the one used by the \textit{index-based} approach.

Therefore, a plan for the \textit{EE-operator}  is a combination of the \textit{index-based} approach and \textit{filter \& verification-based} approach.  Thus the cost of a plan, where $\alpha$ proportion of the entities are processed by the \textit{index-based} approach and $\beta$  proportion are processed by the \textit{filter \& verification-based} approach, is:

\begin{align*}
Cost(plan) = Cost(\alpha, sigX)^{index} + Cost(\beta, sigY)^{ishf\&ssj}
\end{align*}

When only one of the approaches is used, the approach not used contributes a cost of $zero$ to the plan. The signature schemes are denoted by \textit{SigX} and \textit{SigY} for the \textit{index-based} approach and the \textit{filter \& verification-based} approach respectively. Thus searching over the search space is an optimization problem to \textit{minimize} the cost of the plan.

\subsection{Searching the Plan Space}
In this section we describe the algorithm for searching an optimal plan, $\rho$, in a large space. The plan is a function
of the signature scheme and the entity extraction algorithms.s
Suppose that we have three signature schemes \textit{Prefix, Jaccard Variants, and LSH}.
We pick any pair of approaches, for example, \textit{index-based} approach using prefix signatures and  the \textit{filter \& verification-based} using Jaccard Variant signatures and we then seek to determine how to partition the entities. If we do a naive enumeration of the costs  at every possible partitioning point, for a dictionary size of $N$, we do $N$ enumerations. The dictionary is typically large, in the order of millions, thus an exhaustive enumeration would not be practical.

We reduce the number of enumerations by using an efficient search algorithm since the entities are already sorted by occurrence frequency.

\begin{enumerate}
  \item currentCheapestCost = $min(Cost^{index}, Cost^{ishf\&ssj})$
  \item searchRange = 0 - $N$
  \item BinarySearch (searchRange, find new cheapest cost < current cheapest)
  \item currentCheapestCost = newCheapestCost
  \item searchRange = area bounding newCheapestCost
  \item Repeat steps 2-5 over an increasingly narrow search range
  \item Until new cheapest is == current cheapest OR $searchRange$ is 0.
  \item Emit plan(signature, extraction method)
\end{enumerate}

We repeat the procedure for all pairs of approaches. If we have three signature schemes we have maximum of nine
pairs, which is a small constant. For each pair we do binary search over an increasingly small range space,
 $c$ times, which is another small constant. Therefore the complexity of the search algorithm is $\log N$.

We prove that the algorithm works correctly, that is,
its output is the partitioning with the cheapest cost within the joint space signature schemes
and entity extraction algorithms. In order to prove the correctness of the algorithm, we need to show that
both  $Cost^{index}$  and  $Cost^{ishf\&ssj}$  are monotonically non-decreasing functions 
over the sorted space of entities, for any given signature scheme. Since the entities are sorted based on occurrence frequency, based on the
definition 3,  $Cost^{index}$ , the index cost is only affected by the memory size, how many times the entire collection must
be searched is based on the memory capacity and the more the number of the entities the more
times we have to search over the collection, thus index cost is monotonically non-decreasing. Based on definition 4,
the cost of the ssj is based on cost of shuffling, shuffling cost is highest for the most frequently occurring entities
which appear at the beginning since the entities are sorted in descending order of occurrence, thus again
ssj cost is non-decreasing. We thus have the following lemma.

\noindent {\textbf{ LEMMA 1:}}
Given an ordered list of entities in decreasing order of 
Given a collection of entities ordered in descending order
of frequency of occurrence. show this with x, y, like math symbols.

The cost of the filter \& ssjoin approach is defined as:
$Cost^{ishf\&ssj} = \frac{|C|}{|M|}  \times C_{sig(c)}+ |Sig| \left(C_{suffle(sig)} + C_{verify{sig}} \right)$. Where
$C_{sig(c)}$ is the cost of generating signatures, $C_{suffle(sig)}$ is the cost of shuffling signatures and $ C_{verify{sig}}$ is the cost of verifying candidates of a signature.
\\

%Following the LEMMA we have the THEOREM.
%that we get the best thing.
%
%Proof. Binary search will get the best thing if order is monotonically non-increasing.

%\begin{itemize}
%\item Plan space
%
%\item Extended plan space ( + Hybrid approach)
%
%\item Statistics gathering via sampling
%
%\item Searching over plan space (\textit{show algorithm for searching  for plan})
%\end{itemize}

%\section{Choosing execution strategy}
%We summarize what we have achieved with our operator.
%It aims to choose the right execution strategy.
%We give an overview algorithm for doing so, if not already covered in preceding section.

\section{Related work}
%\begin{itemize}
%\item mention  MapReduce cost models  work
%
%\item mention MapReduce Join related work
%
%\item mention work on Text  Analytic optimization.
%\end{itemize}

Chaudhuri et al. \cite{ChaudhuriGK06} eveloped an operator-centric approach
for for set-similarity joins. The main difference in our approach is the target for MapReduce 
algorithms. Sarawagi et al. \cite{ChandelNS06} uses an inverted-index approach to 
compute set overlap string similarities. The main difference with our approach is 
that we do not fix on the index-approach but instead allow the operator to make
cost-based decisions in choosing the best implementation of approcximate entity mentions.

In terms of optimization for text-centric tasks, \cite{IpeirotisAJG06}
introduced an optimizer for choosing between query-based and crawl-based
method for various text-analytics tasks in a cots-based way. The optimizer adaptively selects the best execution
strategy. The EE operator is specifically targeted for MapReduce implementations.

Afrati and Ullman \cite{AfratiU10} investigated the problem
of efficient joins in MapReduce. Vernica et al. \cite{Vernica10} developed
algorithms for set similarty joins in MapReduce. Yang, et al. \cite{Yang07} extended MapReduce to Map-Reduce-Merge, in order to
allow users to express different join types and algorithms. None of these join approaches
propose a cost-model approach for finding the best approach for a given setting.

A number of recent work have studied optimization 
for MapReduce tasks, though none investigate optimization
for approximate mention extraction of entities.
optimization. The Manimal system \cite{Jahani11} analyses MapReduce programs to do 
general database-style optimizations. Dittrich, et al \cite{Dittrich10}. proposed
the use of indices in order to improve MapReduce performance
for certain tasks. 
HadoopDB \cite{Abouzeid09} combines relational and MapReduce qualities into one system.  However, HadoopDB is
designed to be a parallel relational database, it does not optimize MapReduce tasks.

%\input{eeop-experiments}
%\input{eeop-relatedwork}

%\section{Conclusions}

%%%%%%%%%%%%%%%%%%%%%%%%%%%%%%%%%%%%%%%%%%%%%%%%%%%%%%%%%%%%%%%

\end{document}